\pdfoutput=1
\documentclass[aps,prd,tightenlines,floatfix,reprint,twocolumn,groupedaddress,showpacs]{revtex4}
\usepackage{graphicx}
\usepackage{wrapfig}
\usepackage{hyperref}
\usepackage[all]{xy}
\usepackage[usenames]{color}

\newcommand{\nc}{\newcommand}
\nc{\beq}{\begin{equation}}
\nc{\eeq}{\end{equation}}
\nc{\bea}{\begin{eqnarray}}
\nc{\eea}{\end{eqnarray}}
\nc{\n}{\nonumber \\}

\nc {\araa}{Annual Review of Astronomy and Astrophysics}
\nc{\physrep}{Physics Reports}

\nc{\tpm}{\tau^+\tau^-}
\nc{\bb}{b \bar b}
\nc{\ee}{e^+e^-}
\nc{\mm}{\mu^+\mu^-}
\nc{\ww}{W^+W^-}
\nc{\acs}{\ensuremath{ \langle \sigma_{\rm a} v \rangle  }}

\begin{document}  

\date{\today}
\title{Bounds on Dark Matter Properties from Radio Observations of Ursa Major II using the Green Bank Telescope}

\author{Aravind Natarajan}
\email{anat@andrew.cmu.edu}
\affiliation{McWilliams Center for Cosmology, Carnegie Mellon University, Department of Physics, 5000 Forbes Ave., Pittsburgh PA 15213, USA}

\author{Jeffrey B. Peterson}
\affiliation{McWilliams Center for Cosmology, Carnegie Mellon University, Department of Physics, 5000 Forbes Ave., Pittsburgh PA 15213, USA}

\author{Tabitha C. Voytek}
\affiliation{McWilliams Center for Cosmology, Carnegie Mellon University, Department of Physics, 5000 Forbes Ave., Pittsburgh PA 15213, USA}

\author{Kristine Spekkens}
\affiliation{Royal Military College of Canada, Department of Physics, PO Box 17000, Station Forces, Kingston, Ontario, Canada K7K 7B4}

\author{Brian Mason}
\affiliation{National Radio Astronomy Observatory, 520 Edgemont Road, Charlottesville, VA 22903-2475, USA}

\author{James Aguirre}
\affiliation{University of Pennsylvania, Department of Physics and Astronomy, 209 South 33rd Street, Philadelphia, PA 19104, USA}

\author{Beth Willman}
\affiliation{Department of Astronomy, Haverford College, 370 Lancaster Ave, Haverford, PA 19041, USA}

\begin{abstract}
\noindent Radio observations of the Ursa Major II dwarf spheroidal galaxy obtained using the Green Bank Telescope are used to place bounds on WIMP dark matter properties. Dark matter annihilation releases energy in the form of charged particles which emit synchrotron radiation in the magnetic field of the dwarf galaxy. We compute the expected synchrotron radiation intensity from WIMP annihilation to various primary channels. The predicted synchrotron radiation is sensitive to the distribution of dark matter in the halo, the diffusion coefficient $D_0$, the magnetic field strength $B$, the particle mass $m_\chi$, the annihilation rate $\acs$, and the annihilation channel.  Limits on $\acs$, $m_\chi$, $B$, and $D_0$ are obtained for the $e^+ e^-$, $\mu^+\mu^-$, $\tau^+\tau^-$, and $b \bar b$ channels. Constraints on these parameters are sensitive to uncertainties in the measurement of the dark matter density profile. For the best fit halo parameters derived from stellar kinematics, we exclude 10 GeV WIMPs annihilating directly to $e^+e^-$ at the thermal rate $\langle \sigma_{\rm a} v \rangle = 2.18 \times 10^{-26}$ cm$^3/$s at the $2\sigma$ level, for $B >$ 0.6 $\mu$G (1.6 $\mu$G) and $D_0$ = 0.1 (1.0) $\times$ the Milky Way diffusion value. 
\end{abstract}


\maketitle

\section{Introduction}
\noindent Understanding the particle nature of dark matter remains one of the biggest challenges in science today. Weakly Interacting Massive Particles (WIMPs) are among the best motivated candidates for the dark matter of the Universe. WIMPs have weak interactions in addition to gravitational interactions, and therefore thermalize with standard model particles in the early Universe. As the Universe expands, the particles fall out of equilibrium, and the WIMP number density is frozen in. For a typical weak scale interaction rate $\sim$ picobarn $\times$ c, the WIMP density at the present epoch is consistent with the observed value, thus making WIMPs natural candidates for dark matter.

Due to the presence of weak interactions, it is possible to probe dark matter through direct, indirect, and collider search experiments. Interestingly the direct detection experiment DAMA has detected annual modulation in the event rate at the 8.9$\sigma$ level \cite{dama}, and has interpreted this as due to the presence of dark matter in the Galaxy. This very exciting result has received some support from other experiments. Possible evidence for the existence of dark matter has been obtained by the CoGeNT \cite{cogent} and CRESST \cite{cresst} experiments, and more recently, by the CDMS collaboration \cite{cdms}. These experiments favor a light WIMP of mass $m_\chi \sim$ 10 GeV interacting with protons with a spin-independent elastic scattering cross section in the range 0.02 - 0.2 femtobarn. These results however do not seem to agree with the exclusion limits obtained by the XENON-10 \cite{xenon10} and XENON-100 \cite{xenon100} experiments.  

If the dark matter particle has a mass $m_\chi \sim$ 10 GeV and an annihilation rate $\acs \sim$ picobarn $\times$ c, it is possible to probe its properties through the annihilation of particles in high density environments. Particle annihilation at early times may be constrained through precision measurements of the cosmic microwave background \cite{hooper_cmb, slatyer, galli1, galli2, chluba, arvi_cmb, giesen}, and through observations of the Galactic center \cite{finkbeiner511, wmap_haze_1, linden_hooper, gomez_etal}, diffuse gamma ray emission \cite{tavakoli_etal}, and synchrotron emission from the Milky Way \cite{mambrini_etal}. Authors \cite{pier} recently published limits on dark matter properties from radio observations of M31. Constraints on dark matter annihilation from the absence of gamma rays from the dwarf galaxies in the local group have been obtained by \cite{ackermann_etal_for_fermi, koush}, and more recently by \cite{new_fermi}. Observations by the Alpha Magnetic Spectrometer (AMS-02) \cite{ams} of a positron excess may also place competitive bounds on dark matter properties \cite{ams_lim1}. 

Dark matter annihilation results in energy being released in the form of standard model particles, including electrons and positrons that emit synchrotron radiation in a magnetic field. The specific intensity of radiation depends on the energy distribution of the electrons and positrons, which in turn depends on the annihilation channel. Particle annihilation to leptonic states such as $e^+e^-$, $\mu^+\mu^-$, and $\tau^+\tau^-$ results in a large number of electrons and positrons with energies close to the WIMP mass. On the other hand, annihilation to hadronic channels such as $b \bar b$ results in a softer spectrum of electrons and positrons. The computation of branching fractions to various channels is model-dependent. The minimal supersymmetric extension of the standard model (MSSM) for example, favors WIMP annihilation to a $b \bar b$ pair through CP-odd Higgs exchange, for light WIMP masses $m_\chi \sim$ 10 GeV (for a review of WIMP properties, see \cite{susy_jkg}). However, constraints on the Higgs sector from the ATLAS and CMS collaborations \cite{atlas, cms, cms2}, and measurements of the rare decay $B_{\rm s} \rightarrow \mu^+\mu^-$ by the LHCb collaboration \cite{Bs, newbsmumu1, newbsmumu2} have severely constrained light dark matter in the MSSM \cite{arvi_mssm}. In this article, we therefore attempt to be model independent: we will consider many possible primary annihilation channels, and compute the expected synchrotron flux.

Ultra-faint dwarf spheroidal galaxies (dSphs) are promising sources of detectable synchrotron radiation from annihilating dark matter because of their proximity and their inferred high dark matter content \cite{mateo,strigari_etal_2007, strigari_etal_2008, strigari_etal_nature_2008}.  Seven ultra-faint dwarfs (Bo\"{o}tes I, Bo\"{o}tes II, Ursa Major II, Coma Berenices, Willman 1, Segue 1, and Segue 2) lie closer than the nearest low luminosity classical dSphs (Draco and Ursa Minor) \cite{Mcconnachie12}.  Observed line-of-sight velocities of individual stars belonging to the ultra-faint dwarfs suggest that they have significantly higher mass-to-light ratios within their half-radii than the classical dSphs \cite{Geha09}.  This combination of nearby distances and high dark matter densities yield emission measure ($J$) values for the nearest ultra-faint dwarfs up to a factor of ten greater than those for the nearest classical dSphs \cite{ackermann_etal_for_fermi}, making them the most likely places to observe an electromagnetic signature of annihilating dark matter.  Some of the extreme ultra-faint dwarfs ($L \lesssim 10^3 L_{\odot}$) such as Segue 1 have a well measured velocity dispersion, and the appearance of being in dynamical equilibrium \cite{willman11a, simon11a, kirby13}.

Although ultra-faint dwarfs may emit the most observable signatures of annihilating dark matter, the translation of their synchrotron or gamma ray observations into quantitative limits on particle dark matter models is more uncertain than it is for the classical dSphs.  The uncertainties (some of which are difficult to quantify) stem from the small number of stars in the ultra-faints, the uncertainties in their velocity dispersions, and (in some cases) the uncertainties in their dynamical states.  For example, there is a factor of $\sim$2 controversy in the velocity dispersion of Bo\"{o}tes I, owing to subtle differences in the selection of its member stars and in the interpretation of apparent dual kinematic components \cite{munoz06b, koposov11a}.  The small number of stars in the ultra-faints also result in $J$ values that are prior dependent by up to a factor of two, unlike the $J$ values derived from the rich star samples available for classical dSphs \cite{ackermann_etal_for_fermi}.

In previous work \cite{paper1} (hereafter Paper 1), some of us obtained deep 1.4 GHz radio observations of the Draco, Ursa Major II, Coma Berenices, and Willman I dwarf spheroidal galaxies using the Robert C. Byrd Green Bank Telescope (GBT). The radio maps of Ursa Major II and Willman I were used to constrain models considered by \cite{cpu2}. The maps of Draco and Coma Berenices were more difficult to interpret due to the presence of residual foregrounds. In this article, we use the results of Paper I to further explore the nature of dark matter. In Section II, we describe the mechanism of synchrotron radiation from dark matter annihilation. We focus on models of the Ursa Major II data from Paper I: the large observed field provides several independent radial profile data points to work with, and the dark matter content of Ursa Major II is significantly better constrained than that of the  potentially disrupting Willman I \cite{willman11a}. We model the halo of Ursa Major II by a Navarro-Frenk-White (NFW) profile \cite{nfw}, and compute the expected synchrotron radiation for leptonic and hadronic channels. We also consider the effect of uncertainties in the halo profile, and the implications of a core radius on the observed flux. In Section III, we use data from the GBT observations described in Paper I to obtain limits on the annihilation rate $\acs$, the magnetic field strength $B$, the diffusion coefficient $D_0$ and particle mass $m_\chi$, for various primary channels. Finally, we present our conclusions.

\section{Synchrotron radiation from dark matter annihilation}

The stable particles resulting from WIMP annihilation include electrons/positrons, protons/antiprotons, deuterons/antideuterons, neutrinos, and photons. In this article, we will only discuss synchrotron radiation from fast moving electrons and positrons in a magnetic field. The synchrotron flux depends on a number of astrophysical and particle  parameters, which we now explore.

\begin{figure}[!t]
\flushleft
\scalebox{0.74}{\includegraphics{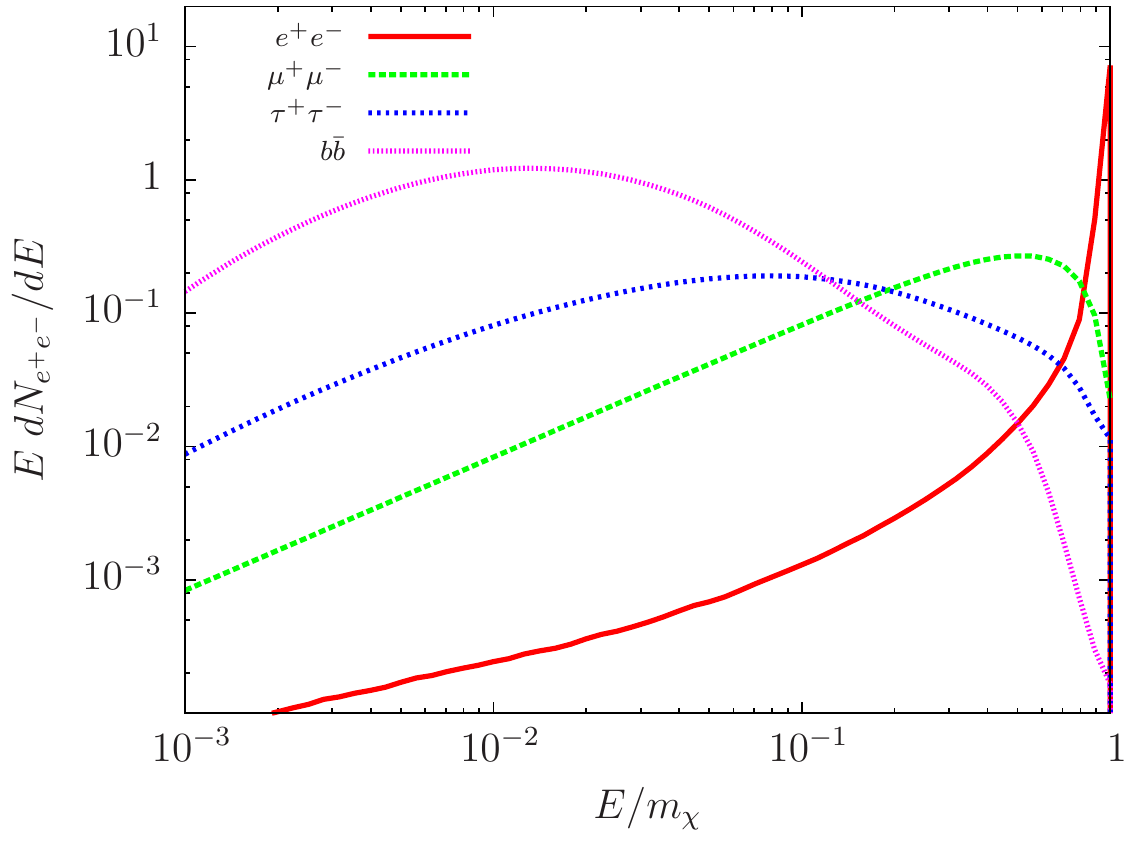}}
\caption{ The dimensionless positron (or electron) spectrum from WIMP annihilation to various primary channels, from \cite{cirelli_etal_2011, cirelli_etal_2011_err, ciafaloni_etal_2011}. For particle annihilation directly to $e^+e^-$, nearly all the energy appears in the forms of electrons and positrons. For other channels 15-30\% of the energy is released in the form of electrons and positrons. Leptonic channels result in a significant number of particles with energy close to $m_\chi$, while hadronic channels produce more particles at lower energies.
\label{fig1} }
\end{figure}

\begin{figure*}[!t]
\begin{center}
\scalebox{0.8}{\includegraphics{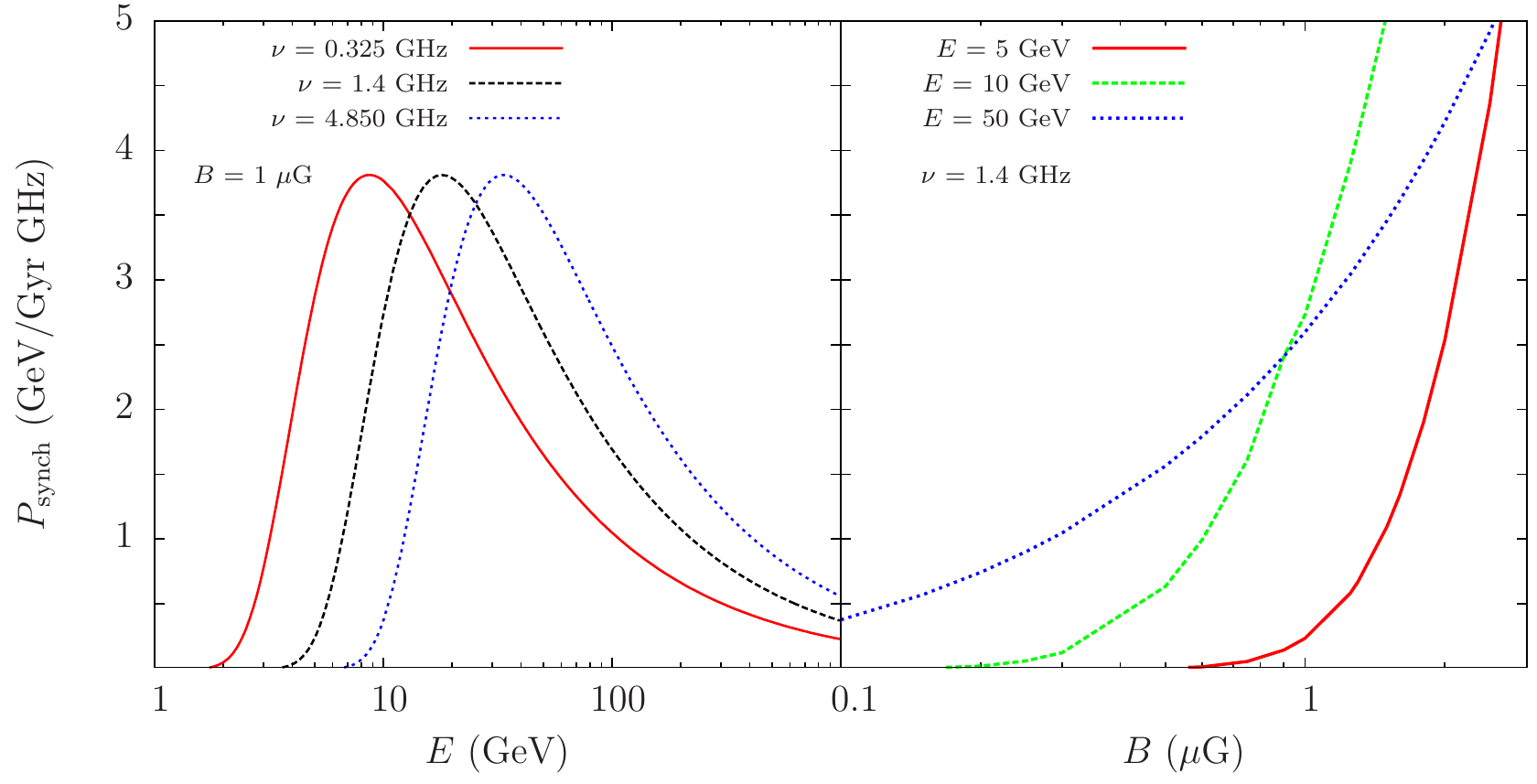}}
\end{center}
\caption{ The synchrotron kernel $P_{\rm synch}$ as a function of particle energy $E$, for different observing frequencies (left panel) and for different values of $B$ (right panel). $P_{\rm synch}$ peaks at $\sim$ 18 GeV for $B = 1 \, \mu$G and $\nu$ = 1.4 GHz.
\label{fig2} }
\end{figure*}

\subsection{Modeling the dark matter halo}

Let us model the dark matter density by a Navarro-Frenk-White profile \cite{nfw}:
\beq
\rho_{\rm DM}(r) =  \frac{\rho_{\rm s}}{x(1+x)^2},
\label{nfw}
\eeq
where $x = r/r_{\rm s}$, and $r_{\rm s}$ and $\rho_{\rm s}$ are constants for a given halo.  Authors \cite{strigari_etal_2007, strigari_etal_2008}  have estimated the values of $r_{\rm s}$ and $\rho_{\rm s}$ from line-of-sight stellar velocities from \cite{simon_geha_2007}, and assuming spherical symmetry and dynamical equilibrium.  The best fit values for $\rho_{\rm s}$ and $r_{\rm s}$ are given by \cite{strigari_etal_2008}:
\bea
\rho_{\rm s} &=& 7.1 \; {\rm GeV}/{\rm cm}^3 \n
r_{\rm s} &=& 0.28 \; {\rm kpc}.
\eea
We discuss the impact of the uncertainties on these values in subsection C.

The dark matter mass within 300 pc is obtained by integrating the density in Eq. \ref{nfw}:
\beq
M(< 300 \, \mathrm{pc}) = 4 \pi \rho_{\rm s} r^3_{\rm s} \; \left [ \ln(1+y) - \frac{y}{1+y} \right ],
\label{mass}
\eeq
where $y =$ 300 pc/$r_{\rm s}$. 
Particle annihilation from dark matter halos is often quantified by means of the emission measure $J$,  which is the dark matter density squared, integrated over the line of sight, and over the resolution of the instrument. Here, we follow the definition of $J$ used by the Fermi collaboration \cite{ackermann_etal_for_fermi}:
\beq
J = \Delta \Omega \left \langle \int_{l.o.s} ds \, \rho^2_{\rm DM}(s) \right \rangle.
\label{eqnJ}
\eeq
$s$ denotes distance along the line of sight, the angle brackets indicate an average over the beam, and $\Delta\Omega \approx \pi  (1.0 \; {\rm deg})^2 / 4$ is the resolution of the Fermi satellite. Although we use the $J$ value as an indicator of annihilation signal strength, we note that the predicted synchrotron emission does not depend directly on the quantity due to diffusion and energy losses (see subsection B).

The distance to Ursa Major II is estimated to be \cite{dallora_etal_2012}:
\beq
L = 34.7^{+0.6+2.0}_{-0.7-1.9} \; {\rm kpc},
\label{uma2_L}
\eeq
where the first error accounts for uncertainties in the calibrated photometry, and the second error accounts for uncertainties in the metallicity of the RR Lyrae star used for calibration.

If dark matter particles have weak interactions, they annihilate at a rate $\acs$, where $\sigma_{\rm a}$ is the annihilation cross section, $v$ is the relative velocity of WIMPs, and the angle brackets denote an average over the momentum distribution. For cold relics, we may expand the annihilation rate in powers of $v$: $\sigma_{\rm a}v = a + bv^2 + \mathcal{O}(v^4)$. For weakly interacting particles, $\langle v^2 \rangle$ at freeze-out $\propto T/m_\chi \approx 1/20$. In the simplest models (unless for e.g. WIMP annihilation is strongly helicity suppressed), the velocity independent term will dominate, and then $\acs$ is nearly independent of velocity. The present day dark matter density fraction obtained by solving the Boltzmann equation takes the value $\Omega_\chi h^2 = 0.11$ when \cite{steigman_etal_2012, arvi_mssm}
\bea
\acs = \acs_0 &=& 2.18 \times 10^{-26} \; {\rm cm}^3/{\rm s} \n
&=& 0.73 \; {\rm pb} \times c.
\eea
We will use $\acs_0$ as our fiducial annihilation rate.

\begin{figure*}[!t]
\begin{center}
\scalebox{0.7}{\includegraphics{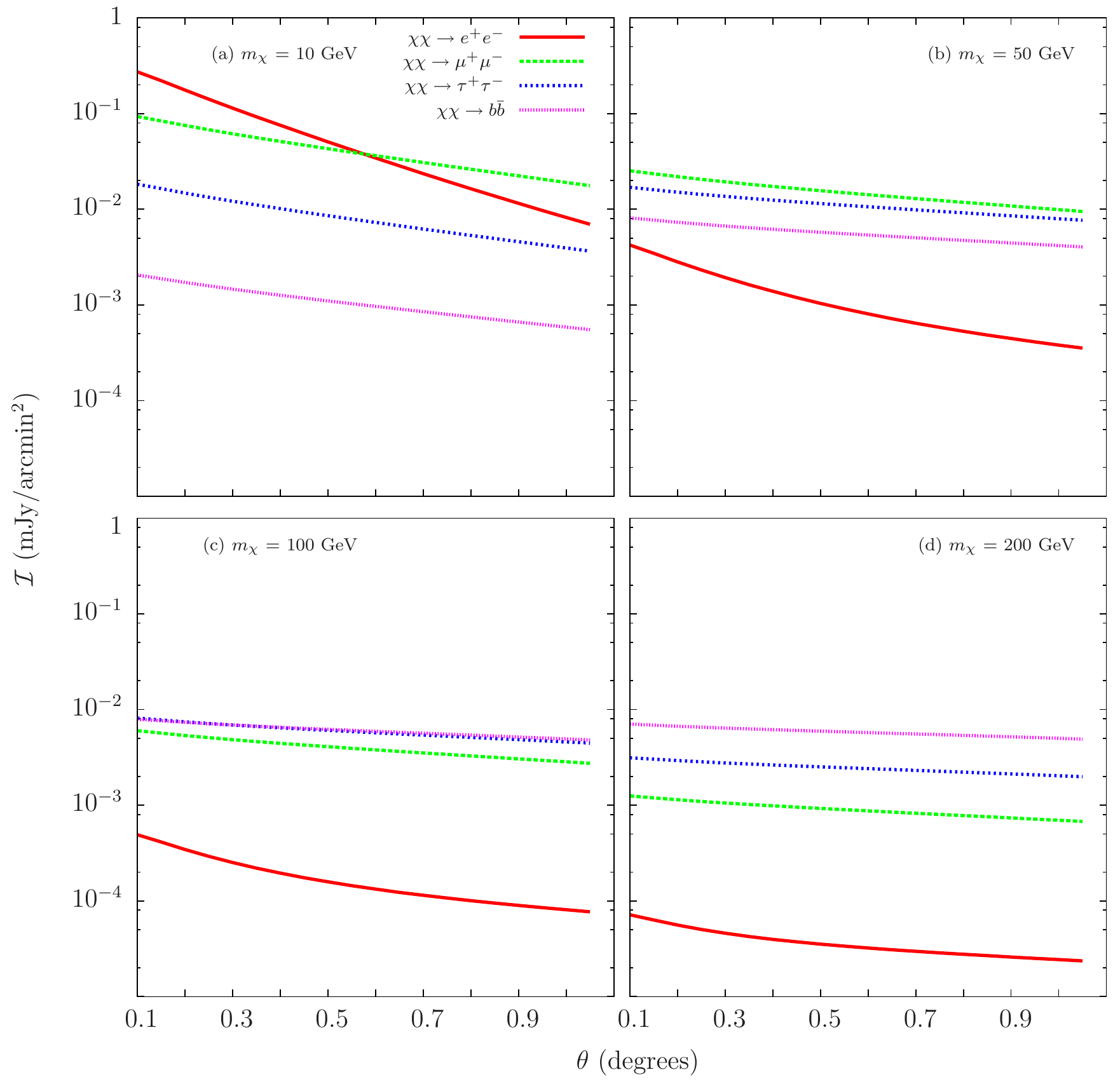}}
\end{center}
\caption{ The predicted specific intensity at $\nu$ = 1.4 GHz, from WIMP annihilation in Ursa Major II, for $D_0 = 10^{-3}$ kpc$^2/$Myr (0.1 $\times$ MW value)  for different values of the particle mass $m_\chi$, and for different annihilation channels. A magnetic field strength $B$ = 1 $\mu$G, and a thermal annihilation cross section $\langle \sigma_{\rm a} v \rangle_0$ are assumed.  For low masses, direct annihilation to $e^+e^-$ predicts the largest synchrotron flux. As the particle mass is increased, we find larger contributions from annihilation to hadronic channels such as $b \bar b$. We note that the synchrotron flux \emph{does not} scale inversely with the particle mass.
\label{fig3} }
\end{figure*}

 Let $n_\chi$ be the number density of dark matter particles in a region of volume $\delta V$. The probability of WIMP annihilation in a time $\delta t$ is then $\acs \delta V \delta t$, while the total number of dark matter particles in the region is $n_\chi \delta V$. The number of annihilations per unit time, per unit volume is therefore $\acs n^2_\chi$. The energy released in electrons and positrons  per unit volume per unit time and per unit particle energy is given by
\beq
Q(r,E) = \frac{ \acs \rho^2_\chi}{m_\chi} \, \frac{dN_{e^+e^-}}{dE}.
\eeq
We will use units of GeV for the particle mass throughout, and GeV/cm$^3$ for the mass density. $dN_{e^+e^-}/dE$ is the number of $e^+e^-$ particles per  energy, normalized to
\beq
\int dE \, \frac{dN_{e^+e^-}}{dE} \leq 1,
\label{sum}
\eeq
since WIMP annihilation typically produces neutrinos and photons in addition to $e^+e^-$ pairs. Fig. \ref{fig1} shows the dimensionless energy spectra of positrons (or electrons) from WIMP annihilation, from \cite{cirelli_etal_2011, cirelli_etal_2011_err, ciafaloni_etal_2011}. Shown are the primary channels $\chi \chi \rightarrow e^+e^-$ (red), $\mu^+\mu^-$ (green), $\tau^+\tau^-$ (blue), and $b \bar b$ (magenta). For direct annihilation to $e^+e^-$, most of the energy remains in the form of electrons and positrons, whereas for other channels, between $15-30$\% of the total energy is released in $e^+e^-$ pairs. Annihilation to leptonic channels results in more particles with high energies compared to hadronic channels.

\subsection{Diffusion of charged particles in a magnetic field}

High energy electrons and positrons produced by particle annihilation diffuse through the magnetic field of the dwarf galaxy, and lose energy through synchrotron radiation and inverse Compton scattering. The number of electrons and positrons per unit volume, per unit energy at a distance $r$ from the center $\psi(r,E)$ is obtained by solving the diffusion equation \cite{cpu1,cpu2}:
\beq
D(E)\nabla^2 \psi(r,E) + \frac{\partial}{\partial E} \left[ b(E) \psi(r,E) \right ] + Q(r,E) = 0
\label{diffusion}
\eeq
$D(E)$ is the diffusion parameter assumed to be independent of position, given by
\beq
D(E) = D_0 \left( \frac{E}{E_0} \right )^\gamma.
\eeq
$D_0$ is the diffusion coefficient, and the index $\gamma$ is set to 0.7, in accordance to the median Milky Way value \cite{donato_etal_2004}. $b(E)$ is the energy loss term due to synchrotron and inverse Compton processes, and takes the form \cite{cpu1}
\beq
b(E) = b_0 \left ( \frac{E}{E_0} \right )^2.
\eeq
$b_0 = 0.788 [ 1 + 0.102 (B/B_0)^2 ]$ GeV/Gyr \cite{cpu1}, where $B_0$ = 1 $\mu$G, $E_0 = 1$ GeV. The energy loss due to Coulomb and Bremsstrahlung processes are  very small for dwarf galaxies for $E > $ 1 GeV \cite{cpu1, cpu2}, and have therefore been neglected.  Under the assumption of stationarity and spherical symmetry, it is possible to solve the diffusion equation (Eq. \ref{diffusion}), and obtain a closed form solution, as done by \cite{cpu1}:
\bea
\psi(r,E) = \frac{1}{b(E)} \int_{E}^{m_\chi} dE' \frac{1}{\sqrt{4\pi\Delta v}} \; \sum_{n=-\infty}^{+\infty} (-1)^n \int_0^{r_h}\frac{r'}{r_n} \n
\times \left [ \exp - \frac{ (r' - r_n)^2}{4 \, \Delta v} - \exp - \frac{ (r' + r_n)^2}{4 \, \Delta v}   \right ]  Q(r', E'). \;\;\;\;
\eea
The diffusion radius $r_{\rm h}$ at which $\psi(r_{\rm h}) = 0$ is assumed to be twice the luminous extent which is $\sim$ 700 pc for Ursa Major II \cite{munoz_etal_2010}, and therefore $r_{\rm h}$ = 1.4 kpc. $r_n = (-1)^n r + 2 n r_h$, and  $\Delta v = v - v'$. $\sqrt{\Delta v}$ is a typical diffusion length, i.e. the average distance traveled by an electron or positron from the point of emission to the point of interaction. $v(E)$ is computed as
\beq
v(E) = \frac{D_0 E_0}{b_0 (1-\gamma)} \; \left [ \left( \frac{E_0}{E} \right )^{1-\gamma} - \left( \frac{E_0}{m_\chi} \right )^{1-\gamma} \right ].
\eeq

Not much is known about diffusion in the ultra faint dwarf galaxies due to their very low luminosity. Authors \cite{rebusco1, rebusco2} studied diffusion of metals by stochastic gas motions in galaxy clusters, and compared their model with observations of peaked iron abundance profiles. They expressed the diffusion coefficient as a product of two terms: $D_0 \propto v \times l$, with $v$ being the characteristic velocity of the stochastic gas motions, and $l$ being their characteristic length scale. From the analysis of a number of clusters, they found that the diffusion coefficient in galaxy clusters is of order $\sim 10^{29}$ cm$^2/$s $\sim 0.3$ kpc$^2/$Myr \cite{rebusco2}. This value of $D_0$ is more than an order of magnitude larger than the corresponding value for the Milky Way $D_{0, {\rm MW}} = 0.01$ kpc$^2/$Myr obtained by \cite{donato_etal_2004} by analyzing the ratio of Boron to Carbon isotopes. Assuming that the diffusion coefficient scales in a similar manner, authors \cite{jeltema_profumo_2008} have suggested that $D_0$ for dwarf galaxies $\sim$ an order of magnitude smaller than that of the Milky Way since the typical virial velocity dispersions associated with the ultra faint dwarfs is at least an order of magnitude smaller than the Milky Way value \cite{simon_geha_2007}. We will therefore choose $D_0 = 10^{-3}$ kpc$^2/$Myr = $0.1 \times  D_{0, {\rm MW}}$ as our fiducial value for the diffusion coefficent, though we also consider the more conservative value $D_0 = D_{0, {\rm MW}}$.

Similarly, not much is known about the turbulent magnetic field strength in the ultra faint dwarf galaxies. Authors \cite{chyzy} performed a search for radio emission in an unbiased sample of 12 local group irregular and dwarf irregular galaxies ($M < 10^9 M_\odot$) with the 100-m Effelsberg telescope at 2.64 GHz and 4.85 GHz. Radio emission from cosmic ray synchrotron was detected for 3 dwarf galaxies with inferred values of the magnetic field equal to 2.8 $\pm$ 0.7 $\mu$G for IC 1613, 4.0 $\pm$ 1.0 $\mu$G for NGC 6822, and 9.7 $\pm$ 2.0 $\mu$G for IC 10, under the assumption of equipartition between the magnetic and cosmic ray energy densities. For comparison, the Small Magellanic Cloud has a known magnetic field $B$ = 3.2 $\pm$ 1.0 $\mu$G. Observations of Ursa Major II and Willman I described in Paper I were used to place a tentative upper limit $B \lesssim$ 1 $\mu$G in equipartition in these galaxies. We will use $B$ = 1 $\mu$G as our fiducial value, but will consider other values of $B$ as well.
\begin{figure}[!t]
\begin{center}
\scalebox{0.75}{\includegraphics{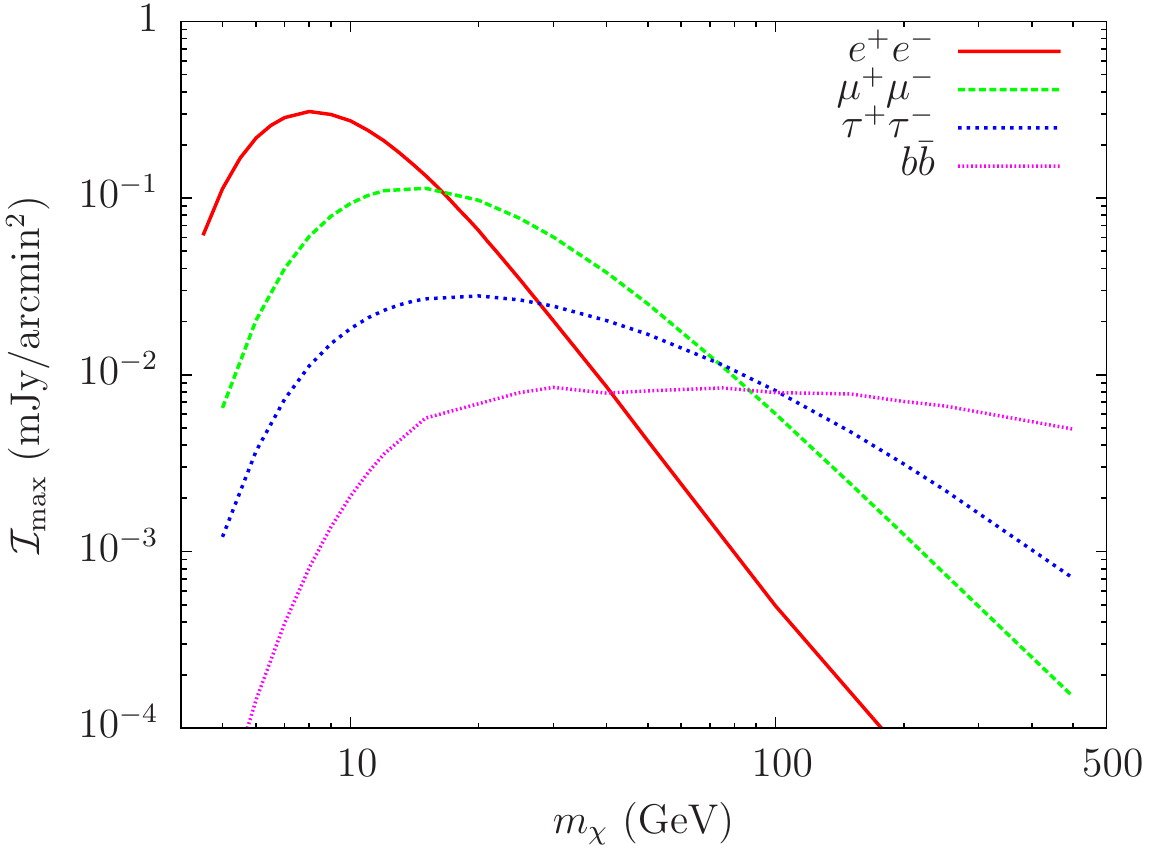}}
\end{center}
\caption{ Variation of synchrotron flux (at $\theta$ = 0.1$^\circ$) with the particle mass. Shown are the channels $\chi\chi \rightarrow e^+e^-$ (red), $\mu^+\mu^-$ (green), $\tau^+\tau^-$ (blue), and $b \bar b$ (magenta).
\label{fig4} }
\end{figure}

The local emissivity $j_{\rm synch}$ (power per unit volume per frequency) is obtained by integrating the electron/positron spectrum over the synchrotron kernel.
\beq
j_{\rm synch} (\nu,r) = \int_{m_{\rm e}}^{m_\chi} dE \,  \frac{dn_{e^+e^-}}{dE} (r,E) \,  P_{\rm synch} (\nu, E),
\eeq
where $dn_{e^+e^-}/dE = [ dn_{e^+}/dE \, + \, dn_{e^-}/dE ] = 2 \times dn_{e^+}/dE$.  The synchrotron kernel $P_{\rm synch}(\nu, E)$ is given by \cite{cpu1, longair}
\bea
P_{\rm synch} &=& \frac{\sqrt{3}}{2} r_0 e B \int_0^\pi d\theta \sin^2\theta \, F \left(\frac{x}{\sin\theta}\right ) \n
&\approx& 2.3 \; \frac{ {\rm GeV} }{ {\rm GHz} \, {\rm Gyr} } \left( \frac{B}{\mu {\rm G}} \right ) \int_0^\pi d\theta \sin^2\theta \, F \left(\frac{x}{\sin\theta}\right ), \;\;\;\; \;\;\;
\eea
where $r_0 = e^2/m_{\rm e}c^2$ is the classical electron radius. $x$ is given by
\beq
x \approx 0.87 \, \left( \frac{ \nu }{1.4 \, {\rm GHz}} \right ) \; \left( \frac{ \mu{\rm G} }{B} \right ) \; \left( \frac{10 \, {\rm GeV}}{E} \right )^2,
\eeq
and $F(t)$ is computed as \cite{cpu1}
\beq
F(t) = t \int_t^\infty dz \, K_{5/3}(z) \approx 1.25 t^{1/3} e^{-t} \left[ 648 + t^2 \right ]^{1/12}.
\eeq
The specific intensity $\mathcal{I}$ may then be calculated by integrating $j_{\rm synch}$ along the line of sight to the dwarf galaxy:
 \beq
\mathcal{I} (\nu) = \frac{1}{4\pi} \; \int_{\rm l.o.s.} ds \; j_{\rm synch}(\nu,s).
\eeq

Fig. \ref{fig2} shows the synchrotron kernel $P_{\rm synch}$ for different frequencies (panel (a)), as well as for different values of $B$ (panel (b)). $P_{\rm synch}$ is very small for energies $E \lesssim$ few GeV, and reaches its peak at energies around 20 GeV for $B$ = 1 $\mu$G, $D_0 = 10^{-3}$ kpc$^2/$Myr, for observing frequency $\nu$ = 1.4 GHz.  Fig. \ref{fig3} shows the predicted specific intensity $\mathcal{I}$ of synchrotron radiation due to WIMP annihilation to various primary channels, for WIMP masses $m_\chi$ = 10, 50, 100, and 200 GeV. For small WIMP masses, direct annihilation to $e^+e^-$ provides the largest signal, which rapidly falls off as the particle mass is increased (see Fig. \ref{fig4}). The contribution from the $b \bar b$ channel on the other hand \emph{increases} as the mass is increased up to $m_\chi \sim$ 200 GeV owing to a larger number of $e^+e^-$ particles in the relevant $10 - 100$ GeV energy range, as seen in Fig. \ref{fig4}. Thus the synchrotron signal does not scale inversely with the WIMP mass, in contrast to gamma ray or CMB observations. The relatively broad energy spectrum for the $b \bar b$ channel results in a slow variation of flux with particle mass (Fig. \ref{fig4}), as well as with angle (Fig. \ref{fig3}).

\begin{figure}[b]
\begin{center}
\scalebox{0.45}{\includegraphics{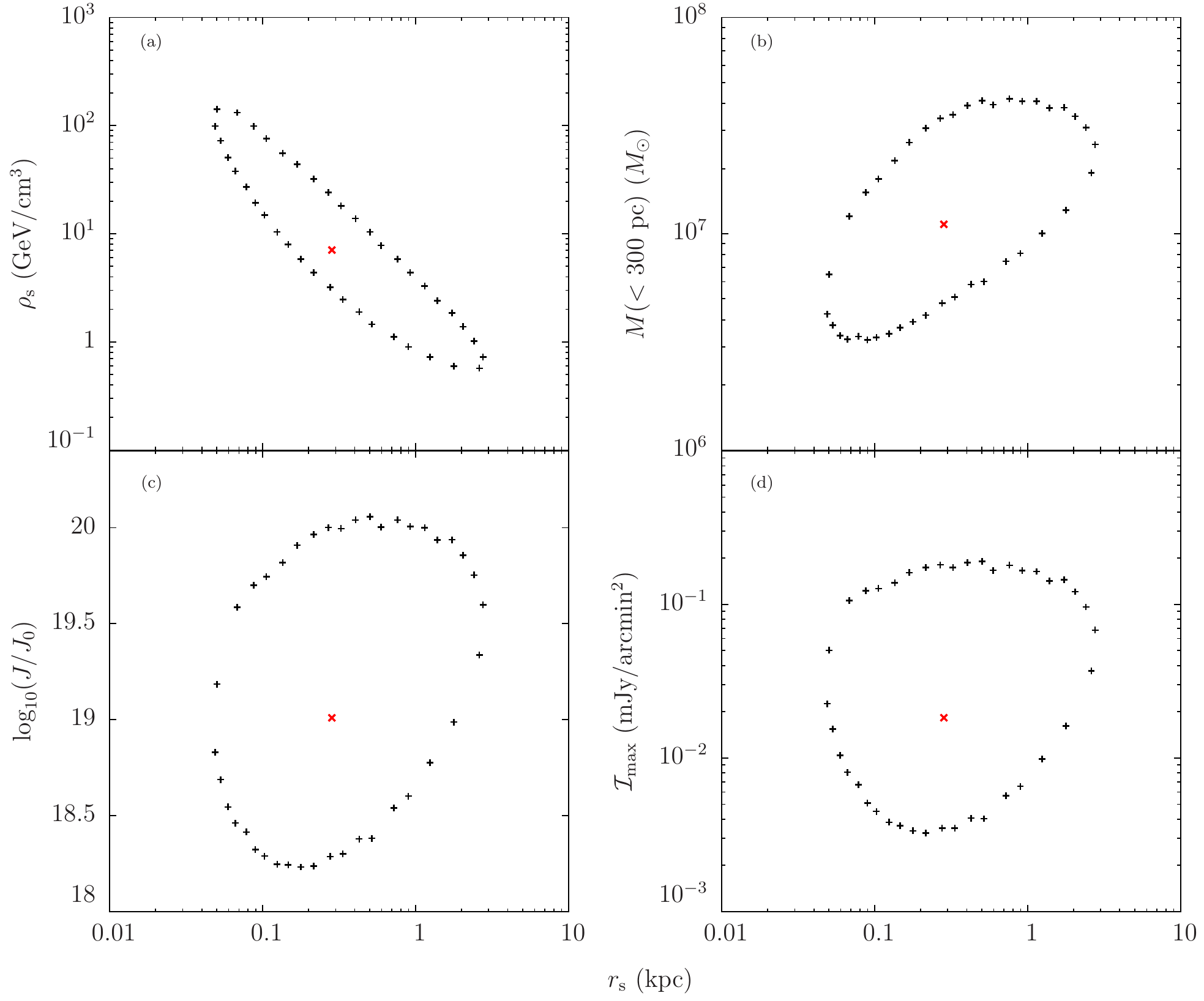}}
\end{center}
\caption{ 
Errors in the determination of halo parameters. Panel (a) shows the 90\% confidence contours in the $r_{\rm s} - \rho_{\rm s}$ plane (from \cite{strigari_etal_2008}), marginalized over velocity anisotropy. Panels (b), (c), and (d) show the corresponding errors in $M(<$ 300 pc), $\log_{10} (J/J_0)$, and $\mathcal{I} (\theta = 0.1^\circ)$ ($\tau^+\tau^-$ channel with $D_0 = 10^{-3}$ kpc$^2/$Myr). The red dots show the best fit parameters.
\label{fig5} }
\end{figure}

\begin{figure*}[t]
\begin{center}
\scalebox{0.75}{\includegraphics{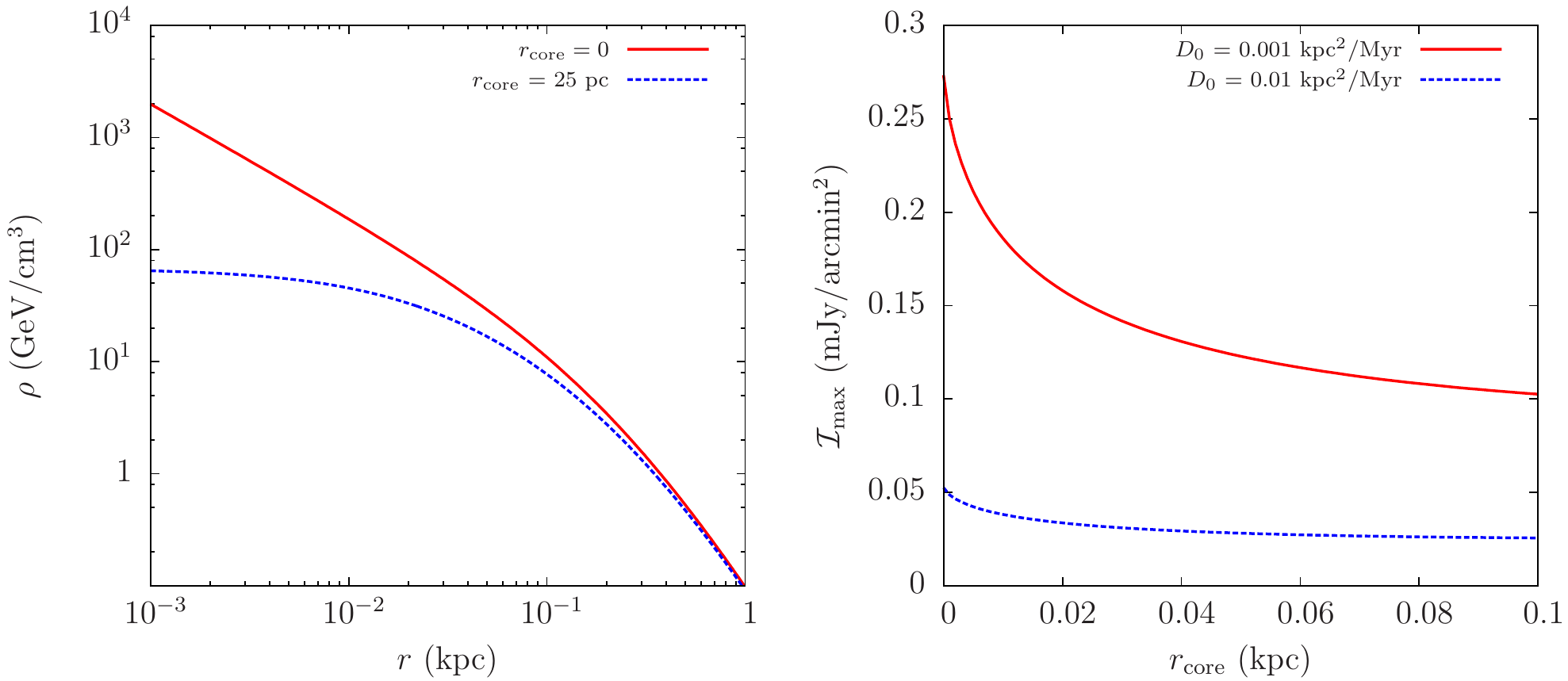}}
\end{center}
\caption{ Effect of a core radius. The left panel shows the NFW density profile (red) along with a profile with non-zero core radius (blue). The panel on the right shows the radiation intensity $\mathcal{I}_{\rm max}$ (at $\theta = 0.1^\circ$) for WIMP annihilation to $e^+e^-$ with $D_0 = 10^{-3}$ (red) and $10^{-2}$ kpc$^2/$Myr (blue) (assuming the thermal rate $\acs_0$). The peak synchrotron flux at $r_{\rm core} = 100$ pc falls to $\sim$ 38\% (49\%) of the value at $r_{\rm core} = 0$ for $D_0 = 10^{-3} (10^{-2})$ kpc$^2/$Myr.
\label{fig6} }
\end{figure*}

\subsection{Importance of halo modeling, and its effect on the predicted synchrotron flux}

So far, we have modeled the dark matter halo by means of an NFW profile with fixed $\rho_{\rm s}$ and $r_{\rm s}$. Let us now account for the uncertainties in these measurements, and examine the effect on the synchrotron intensity. As mentioned earlier, measurements of stellar velocities \cite{simon_geha_2007, strigari_etal_2008} are used to determine the best fit values of parameters $\rho_{\rm s}$ and $r_{\rm s}$. There are however, large uncertainties in the determination of the halo parameters. Fig. \ref{fig5}(a) (from \cite{strigari_etal_2008}) shows the 90\% confidence region in the $\rho_{\rm s} - r_{\rm s}$ plane, marginalized over the velocity anisotropy parameter. The contours shown in Figs. \ref{fig5}(b), (c), and (d) were derived from the constraints shown in Fig. \ref{fig5}(a). Plots (b) and (c) show the corresponding uncertainties in the mass $M(<$ 300 pc), and the emission measure $J$ (where $J_0$ = 1 GeV$^2$/cm$^5$) for the Fermi resolution (see Eqn. \ref{eqnJ}). (d) shows the uncertainty in the predicted synchrotron radiation for WIMP annihilation to $\tau^+\tau^-$ with $D_0 = 10^{-3}$ kpc$^2$/Myr. The red crosses indicate best fit values. As is clear from Fig. \ref{fig5}(d), the predicted synchrotron flux can vary by more than an order of magnitude depending on the halo parameters. Note that the $J$ values corresponding to these values of $\rho_{\rm s}$ and $r_{\rm s}$ are significantly in excess of the range quoted in \cite{ackermann_etal_for_fermi}. Recall that the synchrotron flux is not directly proportional to $J$ due to diffusion and energy losses, in contrast with gamma ray observations.

It is also interesting to ask whether the presence of a core radius could change the predicted synchrotron flux. High resolution rotation curves obtained by The HI Nearby Galaxy Survey (THINGS) show a clear preference for shallower inner profiles in low mass galaxies \cite{things1, things2, things3}: $\rho \sim r^\alpha$, with $\alpha = -0.29 \pm 0.07$, substantially different from the NFW profile. On the other hand, authors \cite{draco} studied the Draco dwarf spheroidal galaxy using the McDonald Observatory VIRUS-W integral field spectrograph, and found excellent agreement with the NFW profile for $r > $ 20 parsec. To test the importance of a possible core radius, we use Eq. \ref{nfw}, but with $x$ replaced by $(r + r_{\rm core}) / r_{\rm s}$. The mass enclosed within 300 parsec is now:
\bea
M (< 300 \, {\rm pc}) = 4 \pi \rho_{\rm s} r^3_{\rm s} \left [ \ln \left( \frac{1+y}{1+x} \right ) \right. \n
\left. - \frac{y}{1+y} (1+x)\left (1-\frac{x}{y} \right ) + x^2 \, \ln \frac{y (1 + x)}{x (1 + y)}   \right ]
\label{mass_core}
\eea
where $x = (r_{\rm core} / r_{\rm s})$, and $y = (300 \, {\rm pc} + r_{\rm core})/r_{\rm s}$. It is easy to check that Eq. \ref{mass_core} reduces to Eq. \ref{mass} as $r_{\rm core} \rightarrow 0$.

Fig. \ref{fig6}(a) shows the modified density profile for $r_{\rm core}$ = 25 pc (blue, dashed curve) compared to the NFW profile (red, solid curve). (b) shows the variation of the predicted synchrotron specific intensity (at $\theta = 0.1^\circ$) as a function of $r_{\rm core}$, when $M(<$ 300 pc) is held fixed to the best fit value. We note that $\mathcal{I}$ varies by about a factor of 2.5 as $r_{\rm core}$ is increased from 0 to 100 pc. This is small compared to the $\sim$ order of magnitude variation in $J$ due to the uncertainty in $\rho_{\rm s}$ and $r_{\rm s}$. We do not consider cored profiles any more in this paper.

\begin{figure*}[t]
\begin{center}
\scalebox{0.8}{\includegraphics{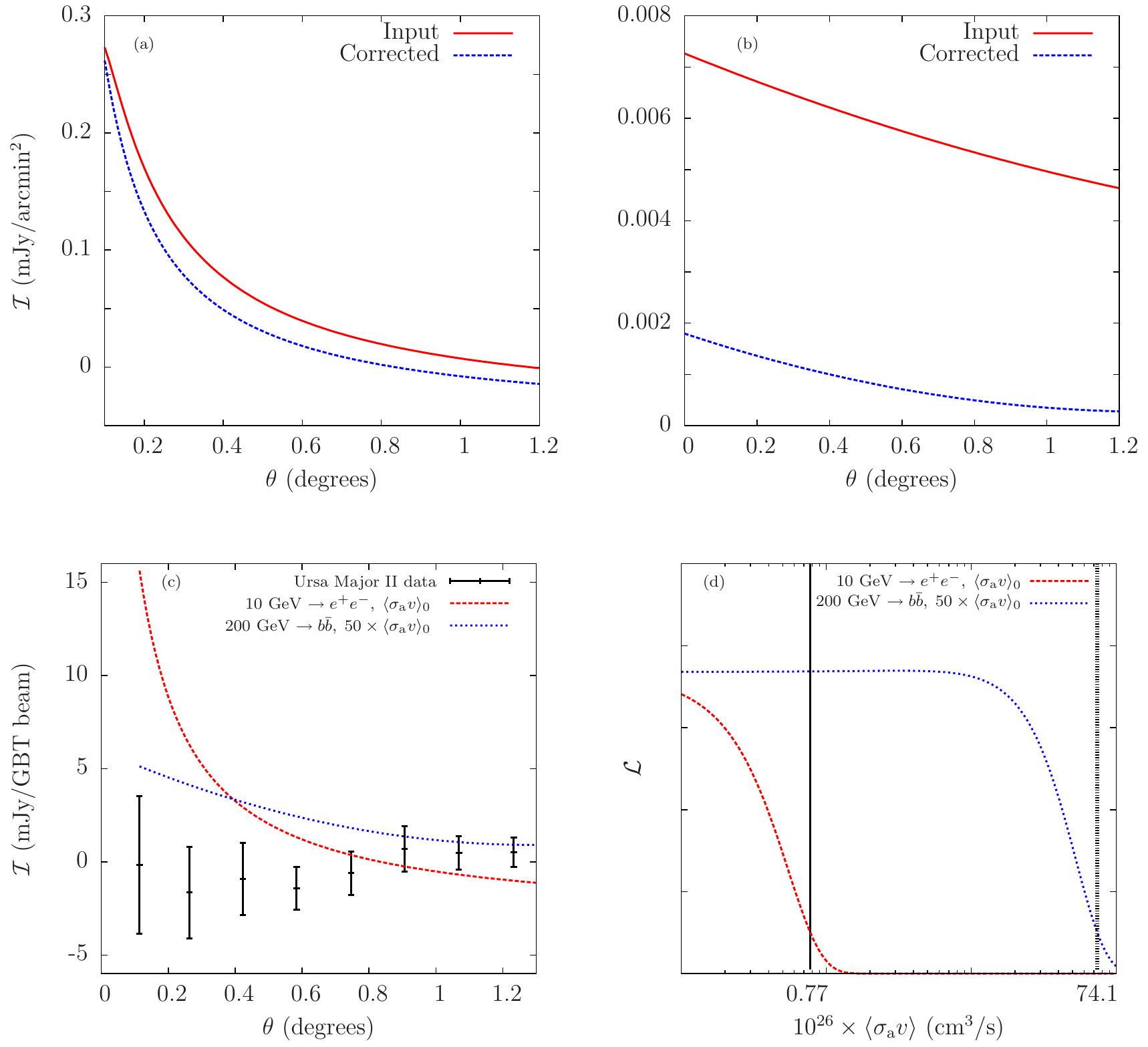}}
\end{center}
\caption{ Linear baselining and comparison with data. Constant emission across the field is indistinguishable from the large Galactic foreground and is removed by the linear baselining procedure. \emph{Top panels:} The solid red curves show the predicted specific intensity of synchrotron radiation $\mathcal{I}$ from Ursa Major II assuming $B$ = 1 $\mu$G and $D_0 = 10^{-3}$ kpc$^2/$Myr, while the dashed blue curves show the intensity accounting for the baselining correction. We consider 2 cases: (i) A 10 GeV WIMP annihilating to $e^+e^-$ (panel (a)) and (ii) a 200 GeV WIMP annihilating to $b \bar b$ (panel (b)). In the first case, much of the radiation is retained due to the rapid fall-off with angle. In the second case however, only 14\% of the flux is transmitted at $\theta$ = 0.5$^\circ$ because of the slow variation of $\mathcal{I}$. \emph{Bottom panels:} (c) shows Ursa Major II data points, from Paper I \cite{paper1}, along with the dark matter contribution (including the baselining correction) for a thermal cross section (10 GeV $\rightarrow e^+e^-$) and 50 $\times$ the thermal cross section (200 GeV $\rightarrow b \bar b$). (d) shows the likelihood function and $2\sigma$ exclusion values for $\acs$, for the two cases considered.
\label{fig7} }
\end{figure*}

Another source of uncertainty is the presence of substructure. Numerical simulations show plenty of substructure in galactic halos. The earliest halos are expected to have formed around $z \sim 60$, with masses $M_{\rm min} \approx 10^{-6} M_\odot$ \cite{green1, green2, small_halos_1,small_halos_2}. Since these halos formed early, they are very compact and may live to the present epoch without significant tidal stripping. If substructure exists in dwarf galaxies, the dark matter annihilation can be substantially boosted. Authors \cite{strigari_etal_2007} find that the boost factor in halos due to substructure can be as large as $\sim 100$ for $M_{\rm min} = 10^{-6} M_\odot$ for a subhalo mass scaling relation $dN/dM \sim M^{-1.9}$, although recent work by \cite{strigari_new} suggests that the boost factor is likely to be small for dwarf galaxies. It is also difficult to place constraints on halo substructure from observations of stellar velocities.  We therefore do not consider a boost factor in our calculations.

\section{Results}

As described in Paper I \cite{paper1}, four dwarf galaxies in the local group namely Draco, Ursa Major II, Coma Berenices, and Willman I, were targeted for observation with the Green Bank Telescope at 1.4 GHz. Of these galaxies, a large map was obtained for Ursa Major II, and it is minimally contaminated by foreground emission. We therefore analyze data from this field to constrain dark matter and astrophysical properties.

\begin{figure*}[t]
\begin{center}
\scalebox{0.8}{\includegraphics{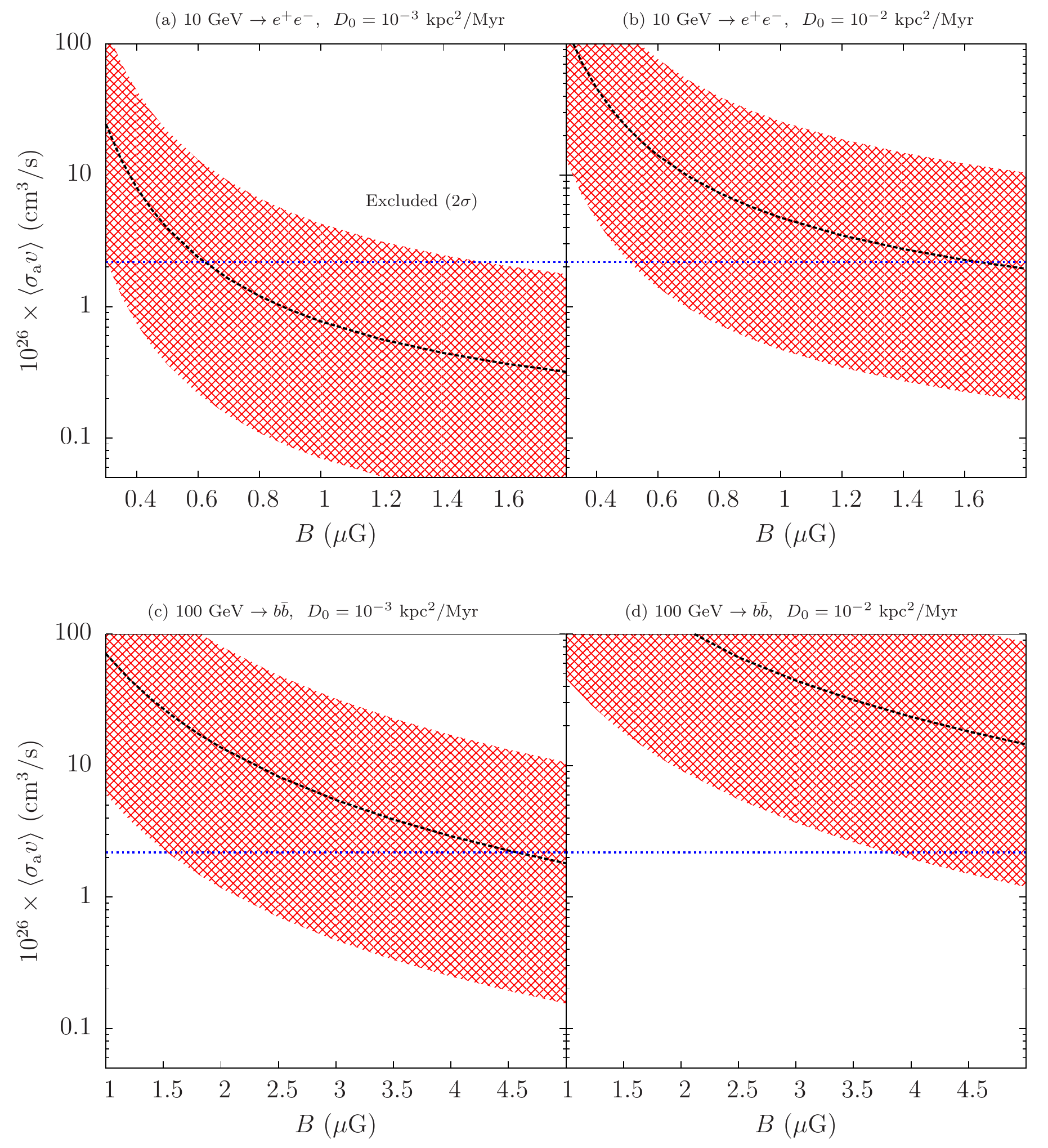}}
\end{center}
\caption{ 2$\sigma$ exclusion curves in the $\acs$ - $B$ plane. We consider the 2 values $D_0 = 10^{-3}, 10^{-2}$ kpc$^2/$Myr, which correspond to approximately, $0.1, 1.0 \, \times$ the median Milky Way value \cite{donato_etal_2004}. The upper 2 panels are for a 10 GeV WIMP annihilating to $e^+e^-$, while the lower panels show a 100 GeV WIMP annihilating to $b \bar b$. The shaded region indicates the uncertainty in the halo parameters (see Fig. \ref{fig5}). The solid black curve is plotted for the best fit halo parameters derived from stellar kinematics. The dashed blue line shows the thermal rate $\acs_0$. For the $e^+e^-$ annihilation channel, we can constrain light WIMP annihilation with a thermal cross section for realistic values of the magnetic field strength $B \sim$ 1 $\mu$G. The $b \bar b$ channel can be probed only for large cross sections or large magnetic fields.
\label{fig8} }
\end{figure*}

The Robert C. Byrd Green Bank Telescope located in West Virginia is a fully steerable, 100m single dish antenna, and is well suited to study radio emission in the 300 MHz - 100 GHz frequency range \cite{gbt}. Source subtraction is achieved using the NRAO VLA Sky Survey (NVSS) \cite{nvss} which is a 1.4 GHz continuum survey covering the entire sky north of $-40^\circ$ declination with a resolution of 45 arcseconds. The availability of NVSS data and the relatively low radio frequency interference (RFI) were the main reasons to choose 1.4 GHz as the observing frequency.  The data is directly calibrated using the NVSS \cite{paper1}:
\beq
d_i = p_i + s \, {\rm NVSS}_i,
\label{calib}
\eeq
where $d_i = d(t_i)$ is the raw GBT time-ordered data, $p_i$ is a first order polynomial to remove baseline drifts in the data, NVSS$_i$ is the NVSS template convolved to the GBT resolution (9.12 arcminutes full width at half maximum), and $s$ is a scale factor determined for each scan. An azimuthally symmetric radial profile centered on the dSph is derived from the calibrated, source-subtracted maps, and uncertainties on the profile points are derived by jackknifing the data. We refer the reader to Paper I \cite{paper1} for more information on the calibration procedure.

A major drawback is our inability to detect constant emission from WIMP annihilation across the galaxy, because it is indistinguishable from the much larger Galactic foreground. The baseline drifts $p_i$ have a characteristic scale on the order of the map size which filters out at least some contribution from dark matter annihilation. In the limit of constant emission across the field, we lose all flux from dark matter annihilation. Conversely, we are fully sensitive to point sources.

\begin{figure*}[t]
\begin{center}
\scalebox{0.55}{\includegraphics{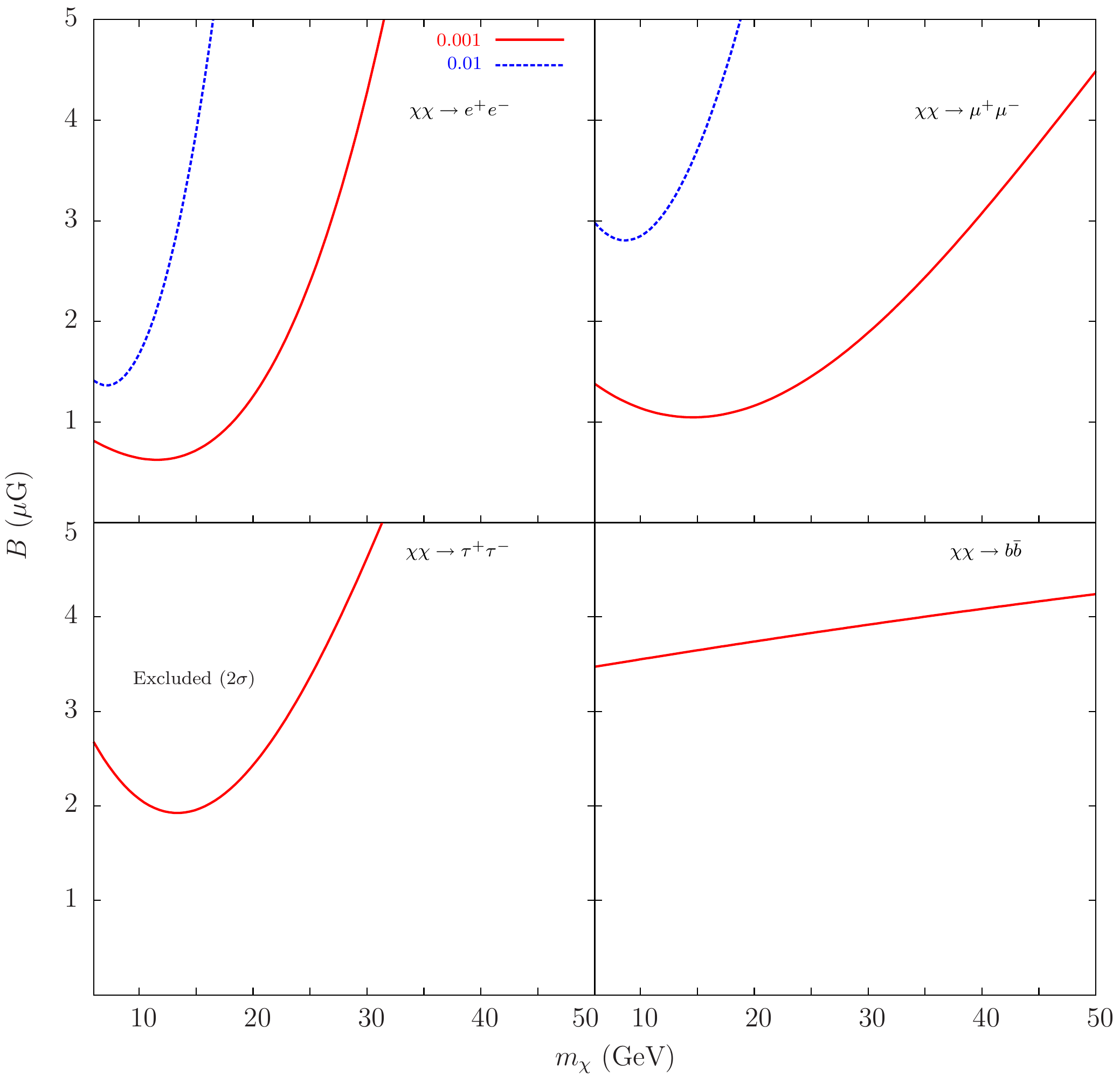}}
\end{center}
\caption{ 2$\sigma$ exclusion curves in the  $B$ - $m_\chi$  plane, for the best fit halo parameters (see Fig. \ref{fig5}), and for a thermal rate $\acs_0$. The solid red curves are plotted for $D_0 = 10^{-3}$ kpc$^2/$Myr, while the dashed blue curves (top panels only) are for $D_0 = 10^{-2}$ kpc$^2/$Myr. Shown are 4 primary channels: $\chi \chi \rightarrow$ $e^+e^-$, $\mu^+\mu^-$, $\tau^+\tau^-$, and $b\bar b$.
\label{fig9} }
\end{figure*}

\begin{figure*}[t]
\begin{center}
\scalebox{0.6}{\includegraphics{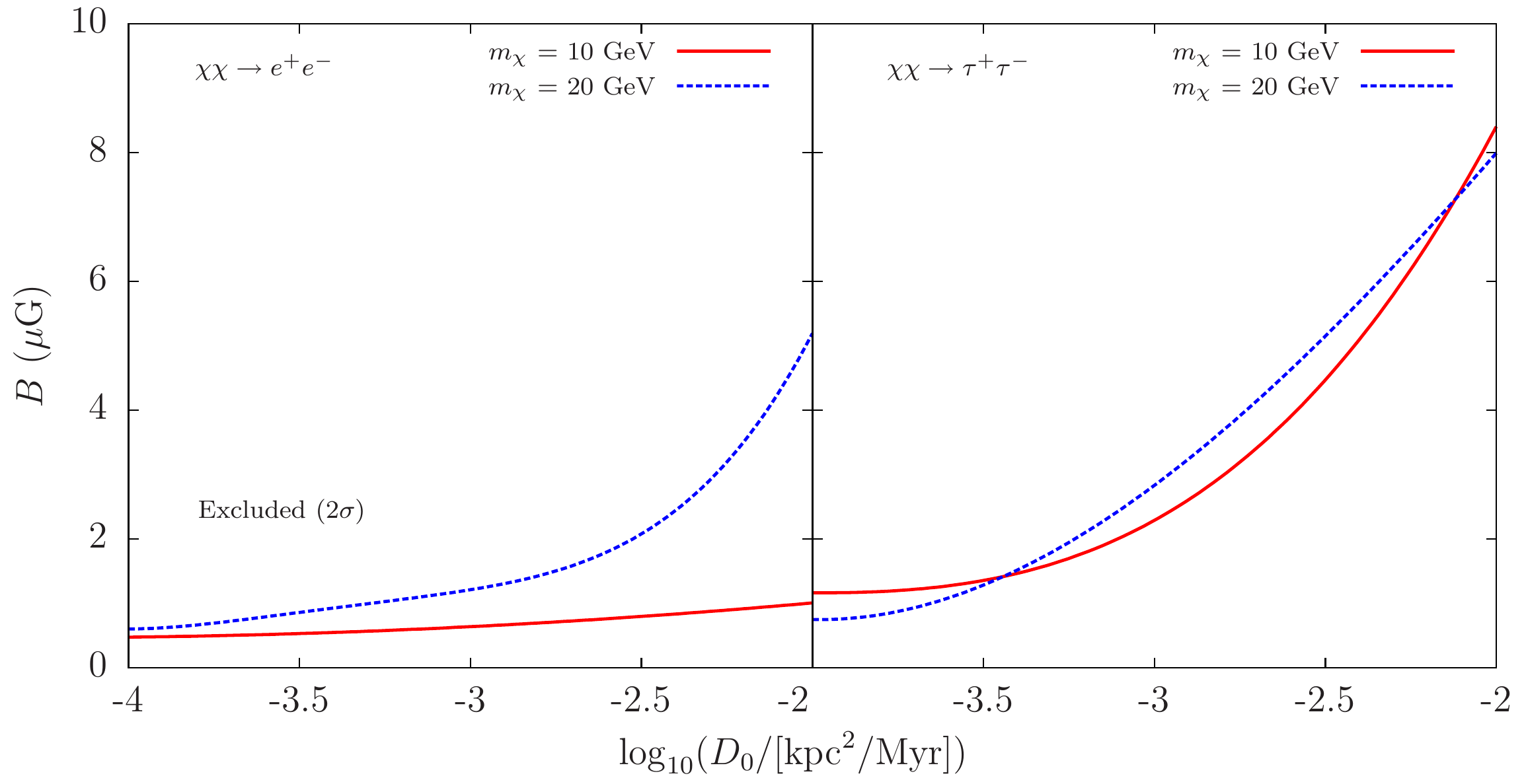}}
\end{center}
\caption{ 2$\sigma$ exclusion curves in the  $B$ - $D_0$  plane, for the best fit halo parameters, and for the thermal rate $\acs_0$. Shown are 2 primary channels: $e^+e^-$ and $\tau^+\tau^-$, for masses $m_\chi$ = 10, 20 GeV.
\label{fig10} }
\end{figure*}

Once the loss of flux due to linear baselining has been accounted for, we may compare our dark matter models with the observations from Paper I \cite{paper1} using the $\chi^2$ statistic defined as:
\beq
\chi^2 = \sum_i \left [ \mathcal{O}(\theta_i) - \mathcal{I}(\theta_i) \right ] \; C^{-1} \; \left [ \mathcal{O}(\theta_i) - \mathcal{I}(\theta_i) \right ]^{\rm T},
\eeq
where $\mathcal{O}(\theta_i)$ denotes the observations at angles $\theta_i$, $C$ is the covariance matrix, and $\mathcal{I}$ has been corrected for flux loss.  The likelihood function $\mathcal{L}$ is constructed from $\chi^2$:
\beq
-2 \ln \mathcal{L} = \chi^2 + {\rm constant}.
\eeq
Fig. \ref{fig7} (a) and (b) show the effect of the linear baselining procedure for 2 cases: (i) a 10 GeV WIMP annihilating to $e^+e^-$, and a 200 GeV WIMP annihilating to $b \bar b$, with $D_0 = 10^{-3}$ kpc$^2/$Myr, and $B$ = 1 $\mu$G. In the first case, we see that the baselining procedure is only a minor correction to the predicted flux. This is due to the rapid fall off of flux with observing angle. On the other hand, it is much harder to observe 200 GeV WIMPs annihilating to $b \bar b$ owing to the slow variation of flux which in turn is due to the much broader energy spectrum from WIMP annihilation. At $\theta = 0.5^\circ$, the baseline corrected flux is $\sim$ 56\% (14\%) of the theoretical prediction for the two cases. Panel (c) shows data points and error bars from Paper I \cite{paper1}, along with the predicted synchrotron signal from dark matter annihilation (including the baselining correction) for $D_0 = 10^{-3}$ kpc$^2/$Myr, and $B$ = 1 $\mu$G. The red curve is for a 10 GeV WIMP annihilating to $e^+e^-$ with the cross section $\acs_0$. The blue curve is plotted for a 200 GeV WIMP annihilating to $b \bar b$ with $\acs = 50 \times \acs_0$. (d) shows the likelihood function for the two cases (fitted to a gaussian), along with the $2\sigma$ exclusion in $\acs$. We see that $\acs \lesssim 0.8 \times 10^{-26}$ cm$^3/$s at the 2$\sigma$ level for $\sim$ 10 GeV WIMPs annihilation to $e^+e^-$, if $B$ = 1 $\mu$G and $D_0 = 10^{-3}$ kpc$^2/$Myr. Bounds on WIMP annihilation to hadronic channels are far weaker due to the broad input energy spectrum (see Fig. \ref{fig1}). We exclude 200 GeV WIMPs annihilating to $b \bar b$ for $\acs \gtrsim 74 \times 10^{-26}$ cm$^3/$s for $B$ = 1 $\mu$G and $D_0 = 10^{-3}$ kpc$^2/$Myr.  This bound is $\sim$ 2 times higher than the exclusion limit obtained by the Fermi collaboration for 200 GeV $\rightarrow b \bar b$ from Ursa Major II \cite{ackermann_etal_for_fermi}.

Fig. \ref{fig8} shows the 2$\sigma$ exclusion curves in the $\acs - B$ plane for $D_0 = 10^{-3}$ and $10^{-2}$ kpc$^2/$Myr ($\approx$ 0.1, 1.0 $\times$ the median Milky Way value \cite{donato_etal_2004}). The solid black curve is plotted for the best fit halo parameters ($\rho_s$ = 7.07 GeV/cm$^3$, $r_s$ = 0.28 kpc). The shaded region represents the uncertainty in the determination of $\rho_{\rm s}$ and $r_{\rm s}$, corresponding to the range of values in Fig. \ref{fig5}(a). The dashed blue line shows the thermal annihilation rate $\acs_0$. The top panels (a) and (b) show 10 GeV WIMPs annihilating to $e^+e^-$, while the bottom panels (c) and (d) show 100 GeV WIMPs annihilating to $b \bar b$.  We see that substantial uncertainties exist in modeling the halo properties from observed stellar velocities, even when assuming an NFW profile.  For panel (a), the thermal cross section is excluded at $2\sigma$ confidence for 10 GeV WIMPs if the magnetic field $B >$ 0.6 $\mu$G, for the best fit halo parameters. However due to uncertainties in the density profile, the $2\sigma$ exclusion may be as low as 0.3 $\mu$G, or as high as 1.5 $\mu$G, for an assumed value of $D_0 = 10^{-3}$ kpc$^2/$Myr.

Fig. \ref{fig9} shows the 2$\sigma$ exclusion curves in the $B - m_\chi$ plane for the best fit halo parameters, and assuming an annihilation rate $\acs_0$.  Shown are limits for the various leptonic annihilation channels $\chi \chi \rightarrow e^+e^-, \mu^+\mu^-, \tau^+\tau^-$ and the hadronic channel $\chi\chi \rightarrow b \bar b$. The red curves are plotted for $D_0 = 10^{-3}$ kpc$^2/$Myr and the blue curves (top panels only) are for $D_0 = 10^{-2}$ kpc$^2/$Myr.  The $e^+e^-$ channel predicts the largest flux for $m_\chi < 20$ GeV, but for larger WIMP masses, the hadronic channel $b\bar b$ may dominate. However, the flux for large WIMP masses is still too small to be observable unless $B > 4$ $\mu$G, for a thermal cross section. Fig. \ref{fig10} shows the 2$\sigma$ exclusion curves in the $B - D_0$ plane for the best fit halo parameters, again for the thermal annihilation rate $\acs_0$. Shown are 2 leptonic channels: $\chi \chi \rightarrow e^+e^-, \tau^+\tau^-$, for WIMP masses $m_\chi$ = 10, 20 GeV. Even for very low values of diffusion, we require magnetic fields in excess of $\sim$ 0.4 $\mu$G to probe dark matter through synchrotron radiation at frequency $\nu$ = 1.4 GHz. We note that if the WIMP mass and cross section may be measured by other experiments, one can obtain interesting results on the astrophysics of dwarf galaxies.

Our error bars are dominated by the mapping stability of the GBT and by foregrounds, not by integration time. We may improve our constraints by obtaining larger maps to better model the foregrounds, and to mitigate the loss of flux when the data is baselined. This should help us improve our constraints for the hadronic channels such as $b \bar b$. Stacking maps of multiple dwarf galaxies will also result in smaller error bars. Improved measurements of stellar velocities can significantly decrease the uncertainties in modeling the dark matter halo of these dwarf galaxies. 

\begin{figure*}[t]
\begin{center}
\scalebox{0.85}{\includegraphics{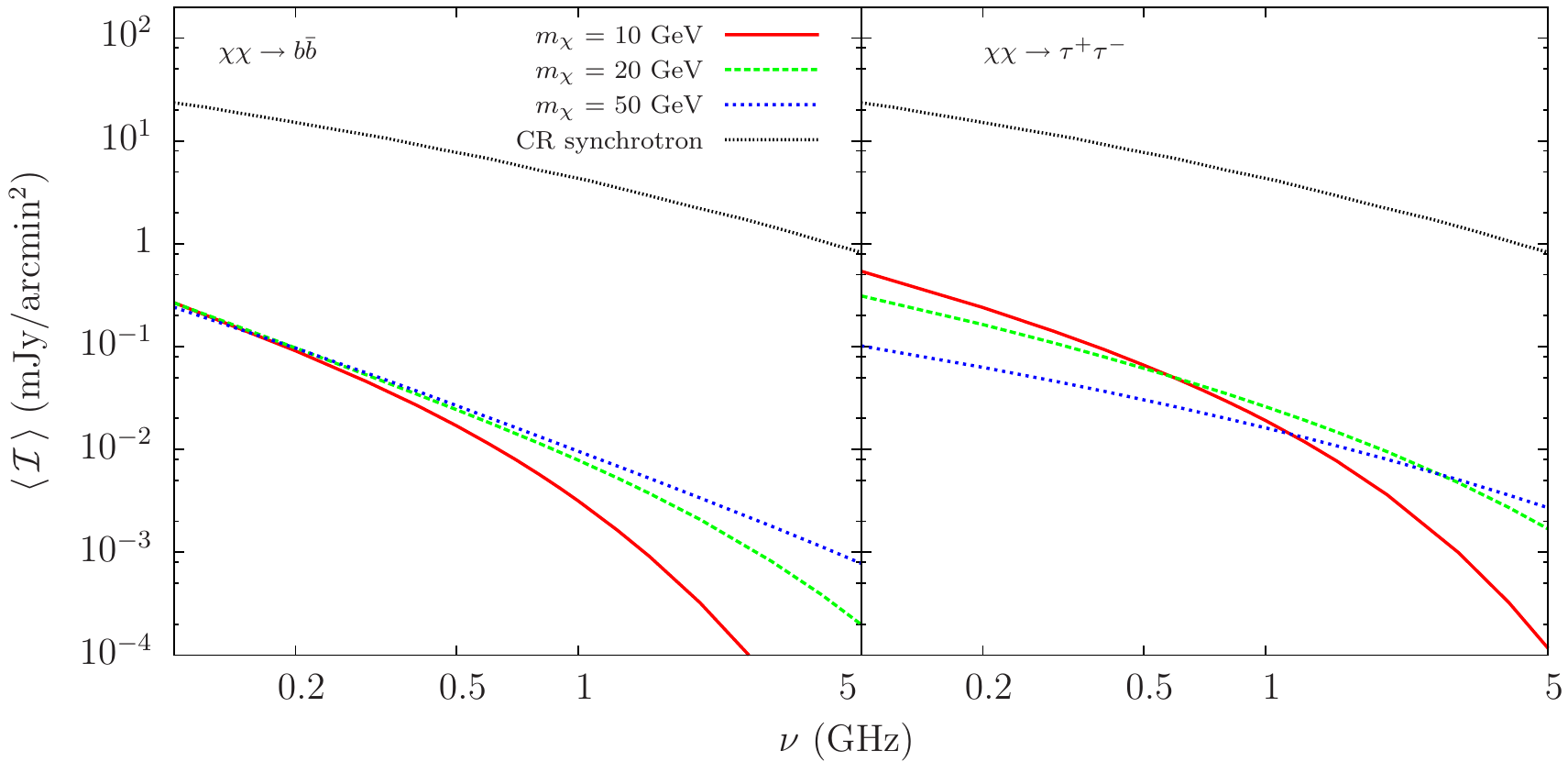}}
\end{center}
\caption{ Specific intensity averaged over a gaussian filter with 1 degree FWHM. Shown is $\langle \, \mathcal{I} \, \rangle$ as a function of observing frequency, for the $b \bar b$ and $\tau^+\tau^-$ channels, for various WIMP masses, for $B $ = 1 $\mu$G and $D_0 = 10^{-3}$ kpc$^2/$Myr. The annihilation rate was set to the thermal value $\acs_0$. The synchrotron spectrum from Galactic cosmic rays is from \cite{strong_etal}.
\label{fig11} }
\end{figure*}

We have only considered observations at 1.4 GHz due to the availability of NVSS point source data, as well as the low RFI contamination at this frequency. However, observations at multiple frequencies can help model and subtract foreground contamination such as synchrotron radiation from cosmic rays from the Milky Way, as well as intra-galactic cosmic rays. Multi-frequency observations are possible using a single dish antenna to map the continuous emission if interferometric observations are simultaneously made to subtract point sources. The Galactic synchrotron emission is however, expected to be significantly larger at lower frequencies. Lower frequencies also suffer from large man-made noise. Fig. \ref{fig11} shows the specific intensity $\mathcal{I}$ at different frequencies, for the $b \bar b$ and $\tau^+\tau^-$ channels, as well as the synchrotron spectrum from cosmic rays. The cross section was set to the thermal value $\acs_0$, with $B$ = 1 $\mu$G, and $D_0 = 10^{-3}$ kpc$^2/$Myr. $\mathcal{I}$ falls off steeply with increase in frequency for low mass dark matter particles due to the energy dependence of the synchrotron kernel (see Fig. \ref{fig2}).  The synchrotron spectrum from cosmic rays although much larger in amplitude, varies differently with frequency \cite{strong_etal} from the predicted synchrotron flux from light WIMPs. Thus by observing at multiple frequencies, one may better identify and subtract foregrounds and obtain better constraints on dark matter properties.

\section{Conclusions}

In this article, we used radio observations of the Ursa Major II dwarf galaxy to place constraints on WIMP dark matter annihilation. We used 18.8 hours of data collected using the Green Bank Telescope (GBT) (described in Paper I \cite{paper1}) at 1.4 GHz. The observing frequency of 1.4 GHz allows us to use data from the NVSS to calibrate, and subtract point sources. The radio frequency interference is also manageable at this frequency at the GBT.

The intensity of synchrotron radiation from WIMP annihilation in dwarf galaxies depends sensitively on both astrophysical and particle physics properties. The astrophysical halo properties we studied include the diffusion coefficient $D_0$, the strength of the magnetic field $B$, and the distribution of dark matter in the halo $\rho(r)$. Due to the low luminosity, no observations currently exist for $D_0$ in the ultra faint dwarfs, while observations of stellar velocities are used to compute the density profile of dark matter. Estimates of $D_0$ have been obtained for galaxy clusters from observations of the distribution of metals, resulting in $D_0 \sim 0.3$ kpc$^2/$Myr for these clusters \cite{rebusco1, rebusco2}. Similarly, the ratio of Boron to Carbon isotopes in the Milky Way has been used to constrain Milky Way parameters \cite{donato_etal_2004} resulting in $D_{0, {\rm MW}} = 0.01$ kpc$^2/$Myr, which is $\sim$ an order of magnitude smaller than the value inferred for clusters. Since the local group dwarf galaxies have virial velocity dispersions that are at least an order of magnitude smaller than the Milky Way value, we may expect $D_0$ to be $\sim 0.1 \times D_{0,{\rm MW}}$ as well. We therefore computed our bounds for $D_0 = 10^{-3}$ kpc$^2/$Myr, but also considered the more conservative value $D_0 = 10^{-2}$ kpc$^2/$Myr. The synchrotron flux also depends sensitively on the magnetic field strength $B$. Authors \cite{chyzy} have studied the brighter dwarf irregular galaxies in the local group, and have found magnetic field strengths $B \sim$ few $\mu$G. The non-detection of synchrotron flux from the ultra faint dwarfs (described in Paper 1 \cite{paper1}) places an upper bound on the equipartition magnetic field strength $B \lesssim$ 1 $\mu$G.

The dark matter distribution in Ursa Major II was modeled by an NFW density profile. The constants $\rho_{\rm s}$ and $r_{\rm s}$, were calculated from observed stellar velocities \cite{strigari_etal_2007,strigari_etal_2008}, with the best fit parameters being  $\rho_{\rm s}$ = 7.1  GeV/cm$^3$ and  $r_{\rm s}$ = 0.28 kpc. However due to the limited number of stars in the ultra faint dwarfs, there are considerable uncertainties in the determination of the halo parameters. We considered the 90\% contour in the $\rho_{\rm s} - r_{\rm s}$ plane from \cite{strigari_etal_2007} and found that the synchrotron flux could vary by an order of magnitude due to these uncertainties. We also considered the possibility of a finite core radius, which decreases the flux by $\sim$ a factor of 2 when $r_{\rm core}$ = 100 parsec. 

We also studied the dependence of synchrotron radiation on particle physics properties namely the particle mass, the annihilation rate, and the annihilation channel. For values $B \sim$ 1 $\mu$G and an observing frequency of 1.4 GHz, the synchrotron power peaks when the electron/positron particle energy is in the 10-20 GeV range. WIMP annihilation to leptonic states such as $e^+e^-$, $\mu^+\mu^-$, and $\tau^+\tau^-$ results in many electrons and positrons being produced at energies close to the particle mass $m_\chi$. We are hence most sensitive to $m_\chi \sim$ 10 GeV WIMPs annihilating to leptons. Annihilation to hadronic channels such as $b \bar b$ results in a much broader energy spectrum for electrons and positrons. The synchrotron power \emph{increases} with increase in particle mass for the $b \bar b$ channel, up to $m_\chi \approx$ 200 GeV. Unfortunately due to the gradual fall off of flux with observing angle, it is challenging to observe WIMP annihilation to $b \bar b$. This difficulty may be mitigated by mapping larger areas around the target dwarf galaxies.

A major difficulty in observing synchrotron radiation from dark matter annihilation is that constant emission across the observing field is indistinguishable from the much larger Galactic foreground. We subtract from our data, a first order polynomial \cite{paper1} to remove baseline drifts, and also to subtract synchrotron radiation from the Milky Way. This process also removes some signal from dark matter annihilation. When the flux falls off rapidly with observing angle as is the case for light WIMPs annihilating to leptonic states, this is a relatively small correction. On the other hand, when the flux decreases only gradually, as in the case of heavy WIMPs annihilating to hadronic channels, we lose nearly all the flux from particle annihilation. We therefore obtain stronger bounds for $m_\chi < 20$ GeV WIMPs annihilating to leptonic states. Larger maps and better understanding of the foregrounds involved will reduce the loss of flux due to linear baselining.

In Fig. \ref{fig8}, we constructed $2\sigma$ contours in the $\acs - B$ plane for $D_0 = 10^{-3}$, $10^{-2}$ kpc$^2/$Myr, showing the uncertainties in the determination of halo parameters. Fig. \ref{fig9} shows the $2\sigma$ contours in the $B - m_\chi$ for the best fit halo parameters, for fixed cross section $\acs_0$, for various annihilation channels. For our fiducial parameter choices $B$ = 1 $\mu$G and $D_0 = 10^{-3}$ kpc$^2/$Myr, we exclude 10 GeV WIMPs annihilating directly to $e^+e^-$ for a cross section $\acs > 0.77 \times 10^{-26}$ cm$^3/$s at the $2\sigma$ level. Conversely, for a fixed annihilation rate $\acs_0$, we exclude magnetic field strengths $B >$ 0.6 $\mu$G.  Fig. \ref{fig10} shows the 2$\sigma$ exclusion curves in the $B - D_0$ plane for the best fit halo parameters, again for $\acs_0$, for WIMP annihilation to  $e^+e^-$ and $\tau^+\tau^-$, for $m_\chi$ = 10, 20 GeV. We see that a magnetic field strength of at least $\sim$ 0.4 $\mu$G are required to probe dark matter annihilation with the thermal cross section, even for very low diffusion. In Fig. \ref{fig11}, we computed the specific intensity of radiation due to dark matter annihilation for different observing frequencies, and compared the values with the Galactic cosmic ray synchrotron spectrum.

It is interesting to compare our constraints on the WIMP mass with the results from other experiments. As mentioned earlier, the CDMS collaboration \cite{cdms} recently announced results consistent with the presence of dark matter in the halo of the Milky Way, with a mass of 8.6 GeV, although this result may be in disagreement with other direct detection experiments such as XENON-10 \cite{xenon10} and XENON-100 \cite{xenon100}.  Our results from synchrotron observations of Ursa Major II are in conflict with CDMS if WIMPs annihilate entirely to $e^+e^-$ with a thermal cross section, for our fiducial values $B$ = 1 $\mu$G and $D_0$ = 0.1 $\times$ the Milky Way value. Other annihilation channels and/or lower magnetic fields and larger diffusion values are less constraining. We also note that dark matter masses as low as those preferred by the DAMA \cite{dama}, CoGeNT \cite{cogent},  CRESST \cite{cresst}, and CDMS \cite{cdms} experiments may run into difficulty with CMB \cite{hooper_cmb, slatyer, galli1, galli2, chluba, arvi_cmb, giesen}, Fermi \cite{ackermann_etal_for_fermi, koush}, and AMS-02 \cite{ams_lim1} observations as mentioned in the Introduction. Strong constraints on $m_\chi < 30$ GeV WIMPs have been obtained from LHC data through observations of the Higgs sector \cite{arvi_mssm}, although these bounds are specific to the MSSM.  Our radio observations may be used to place useful constraints on halo properties if the dark matter properties can be measured by other experiments. Larger WIMP masses, i.e. $m_\chi > 50$ GeV may be probed by radio observations at multiple frequencies (as seen in Fig. \ref{fig11}, 50 GeV dark matter predicts a greater flux at frequency $\nu$ = 5 GHz compared to the $m_\chi$ = 20 and 10 GeV cases). Thus synchrotron measurements are complementary to information obtained from other means.

The synchrotron radiation from dark matter annihilation may be substantially increased if significant substructure exists in them. While plenty of substructure is expected from numerical simulations, it is difficult to place any observational constraints on sub-halos within dwarf galaxies. From calculations of the kinetic decoupling temperature for WIMPs, the earliest halos are expected to be $\sim$ few earth masses, and would have formed at $z \sim$ 60 \cite{green1, green2,small_halos_1,small_halos_2}. If these compact planetary mass halos survive to the present epoch, the annihilation rate may be boosted \cite{strigari_etal_2007}, although the effect is not likely to be very large for dwarf galaxies \cite{strigari_new}. Better modeling of dwarf galaxies through high resolution numerical simulations is required to address this issue. A second mechanism that might increase the synchrotron flux is the possibility of Sommerfeld enhancement in the annihilation rate. If dark matter particles annihilate through a light mediator, then the annihilation rate could be enhanced when the WIMP relative velocity is small, as is the case for dwarf galaxies \cite{nima,bovy,robertson_zentner_2009,essig_etal_2009}. We leave a detailed study of these effects to future work.

\acknowledgments{A.N., J.B.P., and T.C.V. acknowledge funding from NSF grant AST-1009615. A.N. thanks the Bruce and Astrid McWilliams Center for Cosmology for partial financial support. K.S. acknowledges support from the Natural Sciences and Engineering Research Council of Canada. B.W. acknowledges funding from NSF grants AST-1151462 and AST-0908193. The National Radio Astronomy Observatory is a facility of the National Science Foundation operated under cooperative agreement by Associated Universities, Inc.}

\bibliographystyle{revtex}
\bibliography{references}

\end{document}